\documentclass[conference, a4paper]{IEEEtran}
\IEEEoverridecommandlockouts
\usepackage{cite}
\usepackage{amsmath,amssymb,amsfonts}
\usepackage{algorithmic}
\usepackage{graphicx}
\usepackage{flushend}
\usepackage{textcomp}
\usepackage{xcolor}
\usepackage{dirtytalk}
\usepackage[autostyle]{csquotes}
\def\BibTeX{{\rm B\kern-.05em{\sc i\kern-.025em b}\kern-.08em
    T\kern-.1667em\lower.7ex\hbox{E}\kern-.125emX}}

\begin{document}
\title{Video Analytics in Elite Soccer: A Distributed Computing Perspective}

\author{\IEEEauthorblockN{Debesh Jha\IEEEauthorrefmark{1}, Ashish Rauniyar\IEEEauthorrefmark{2}, H{\aa}vard D. Johansen\IEEEauthorrefmark{4}, Dag Johansen\IEEEauthorrefmark{4},\\ Michael A. Riegler\IEEEauthorrefmark{3}\IEEEauthorrefmark{4}, P{\aa}l Halvorsen\IEEEauthorrefmark{3}\IEEEauthorrefmark{5}, Ulas Bagci\IEEEauthorrefmark{1}
}
\IEEEauthorblockA{\IEEEauthorrefmark{1}Northwestern University, USA,\\ 
\IEEEauthorrefmark{2}SINTEF Digital, Norway,\\ 
\IEEEauthorrefmark{3}SimulaMet, Norway,\\
\IEEEauthorrefmark{4}UiT The Arctic University of Norway, Norway,\\
\IEEEauthorrefmark{5}Oslo Metropolitan University, Norway\\
Email: debesh.jha@northwestern.edu}}
\maketitle

\begin{abstract}
Ubiquitous sensors and Internet of Things (IoT) technologies have revolutionized the sports industry, providing new methodologies for planning, effective coordination of training, and match analysis post game. New methods, including machine learning, image and video processing, have been developed for performance evaluation, allowing the analyst to track the performance of a player in real-time. Following FIFA's 2015 approval of electronics performance and tracking system during games, performance data of a single player or the entire team is allowed to be collected using GPS-based wearables. Data from practice sessions outside the sporting arena is being collected in greater numbers than ever before. Realizing the significance of data in professional soccer, this paper presents video analytics, examines recent state-of-the-art literature in elite soccer, and summarizes existing real-time video analytics algorithms. We also discuss real-time crowdsourcing of the obtained data, tactical and technical performance, distributed computing and its importance in video analytics and propose a future research perspective.


\end{abstract}

\section{Introduction}
Technology has a vast impact on sport industry, in particular soccer~\cite{andreassen2019real}. Advancements in online connectivity and the rapid proliferation of smartphones and social media websites has brought fans closer to the action than ever before. Spectators now have more alternatives of equipment and experience than ever before~\cite{gantz2014sports}. Elite sports clubs are continually looking for new methods and strategies that can provide them with a competitive edge over their competitors. Thus, the rise of real-time sports data analytics in favourite sports such as basketball, soccer, field hockey, and baseball has changed the nature of sports science. This science is driven by the advanced technique to gather a massive amount of data while the match is in progress. Metrics related to game events like passes, shots, and tackles are gathered carefully. Using acquired data to gain a competitive advantage, such as supporting players in real-time or during training, recruitment, and preparation, is a major challenge these days.  

Today, most professional sports teams have either an analytics department or data scientists~\cite{steinberg2015changing}. Teams frequently scan scout notes from clipboards, gather video-based or sensor information and hand over those files to the elite data scientists. This information is sent to the mathematicians who analyzes the results produced by the programmer. The results even help to decide which player fits the best in their club. This is why analytics is now touted as the ``present and future of sports''~\cite{steinberg2015changing}. Any team that does not fully utilize these will be at a competitive disadvantage. The rise in popularity of data-driven prediction in sports has energized fans, who are consuming more analytical content than ever before. For example, various websites are devoted to the analysis and investigation of sports statistics and how they relate simply to make a prediction. As a result, developing data culture within a sports club's organization is critical for sustaining a competitive edge in today's extremely demanding digital world.

On the other hand, crowdsourcing technology has scaled up the fans' engagement beyond 90 minutes of the match. Crowdsourcing technology is now used by the clubs by broadcasting certain tasks and asking followers to provide input. This feedback is utilized to help solve specific problems. They use the method to reaffirm fans' experiences, which aids in the development of trust among audiences and clubs. Despite the fact that there are few works and case studies on sports analytics, they are either only focused on real-time analytical systems or on predictive analytics~\cite{stensland2014bagadus,chyronhego1} or position of the specific player~\cite{vigh2018position} or post-match monitoring fatigue in professional soccer~\cite{carling2018monitoring,johansen2020scalable}. The current study intends to highlight real-time analysis of the integrated system, motion analysis, crowdsourcing, tactical and technical performance, individual subjective report, and the role of distributed computing in sport analytic.

The main contributions of this paper are: (i) we provide examples of some of the state of the art video analytics, (ii) we describe the connection between elite soccer analysis and distributed computing; and finally, (iii) we provide future perspectives and directions in the field.

\section{Literature review}
The sport industry spends an enormous amount of resources for performance analysis of their games. They use both manual and analytical tools, enabling the team managers and trainers to analyze the game to boost performance. For example, in Interplay-sports~\cite{INTERPLAYSPORTS}, video streams are analyzed and annotated manually using soccer ontology classification strategy. Prozone~\cite{PROZONE} uses video-analysis software that automates some of the manual annotation processes. Specifically, it quantifies players' gestures such as speed, velocity, and position during the game. The measurement of such characteristics has already proven successful in Old Trafford in Manchester and Reebok Stadium in Bolton (now called as University of Bolton Stadium)~\cite{valter2006validation}. Likewise, STATS SportVU Tracking Technology~\cite{STATS} utilizes video cameras to accumulate the positional data of those players over the playing field in real-time. These data are used to improve the players' performance. Although Camargus~\cite{CAMARGUS} produces a superior video technology infrastructure, it lacks other analytical applications. The TRACK~\cite{b} and ZXY Sport Tracking~\cite{chyronhego1} systems use the global positioning system and radio-based systems for capturing performance measurements in athletes. The player's statistics can be presented along with the fitness graph, speed profiles and accumulated distances in many ways like charts and animations. Osgnach et al.~\cite{osgnach2010energy} proposed a match analysis strategy by performing a comprehensive evaluation of soccer players' metabolic requirements by video match analysis to consider accelerations.   

The Muithu system~\cite{johansen2012muithu} merges coach annotations with associated video sequences. However, the video has to be manually transferred and mapped to the game's timeline. Halvorsen et al.~\cite{halvorsen2013bagadus} and Saegrov et al.~\cite{saegrov2012bagadus} demonstrated a system named Bagadus. This system combines a camera array video capture platform with the ZXY Sport Tracking system for participants' statistics along with a method for human expert annotations. Mortensen et al.~\cite{mortensen2014automatic} introduced the automatic video extraction capabilities of Bagadus. Barros et al.~\cite{barros2018distributed} showed a system based on the use of distributed mobile devices that allows the annotation of soccer matches in real-time or after the game is finished (by the observation from another media). In this paper, we cover the basics of sports analytics system to the application of distributed fog computing and IoT to sports analytics. 

\section{Methods}
Match analysis can be used to assess the physical abilities of skilled soccer players, particularly in identifying high intensities, which are also considered as fast running speeds. It is divided into three categories: technical (skill performance), tactical or strategic, and physical. The systems that provide a solution to sport analytics are described below.

\subsection{Motion Analysis}
Motion analysis is the most extensively utilized technique in sports bio-mechanics and rehabilitation for individual player analysis, emphasizing movement patterns and ground activity. The essential information such as total distance travelled, the time taken to complete a specific activity, and efforts applied during varying movement categories, for instance, walking, jogging, standing, sprinting are collected~\cite{data2008role,dwyer2012global}. This information is utilized to create an extensive players activity profile~\cite{bradley2010high}, defining the average physical requirements of each participant along with their playing position on the ground. These obtained data points assist data scientists, trainers, and other professionals in monitoring fluctuations in physical performance over time, allowing them to quantify players' training burden or compare players with similar features or attributes. Furthermore, this information allows the player's activity profile to be correlated to a similar demographic (e.g., teammate or opponent).

\subsection{Tracking using LPM (Radio Signals) and GPS in a Professional Football Club}
The Local Position Measurement (LPM) system is one of the most accurate sports tracking systems. This system uses high-tech RFID technology. LPM system consists of both base stations (antenna) and transponders. Base stations are strategically placed around the ground to collect data. The players wear transponders. The position of the transponders is quantified by the base station, which means that the positions of the players are likewise quantified. The positions are calculated in real time. As a result, the measured data could even be examined while the measurement was taking place. These positioned data are processed to separate standard values like as distance covered, speed and acceleration, and so on. In GPS, the devices are passive receivers of signals from aerial satellites, but LPM systems use wearable emit signals that are relayed to local receivers, which perform the specific triangulation.

\subsection{ZXY Sports Tracking}
The ZXY Tracking system is a commercial product by ChyronHego~\cite{CHYRONHEGO}. ZXY Arena wearable tracking employs RF-based technology and is designed for permanent installations such as specialized training facilities or contest arenas. This entire system is currently positioned at Alfheim arena. The system contains eleven stationary radio receivers mounted around the field. Each of the receivers has a nearly 90-degree field of view. This forms an overlapping zone in the soccer field, which provides high immunity to indicating signal blocking and occlusion. For signal transmission and radio communication, the installed system is dependent on the 2.45\,GHz ISM band. Each radio receiver calculates the positional data depending upon the radio signals obtained. Wearable belts with sensors that the players wear during the game collect the radio signal. This belt has a compass, a heart-rate monitoring system, a gyroscope, and accelerometers. All of these instruments work together to produce positioning data for the players on the field as well as performance measures. Using a sensor, the ZXY system determines the direction of the players on the field, heart rate frequency, location, and step frequency with a sample rate of around 20 times per second.

\subsection{Bagadus}
\begin{figure}
    \centering
    \includegraphics[scale=0.32]{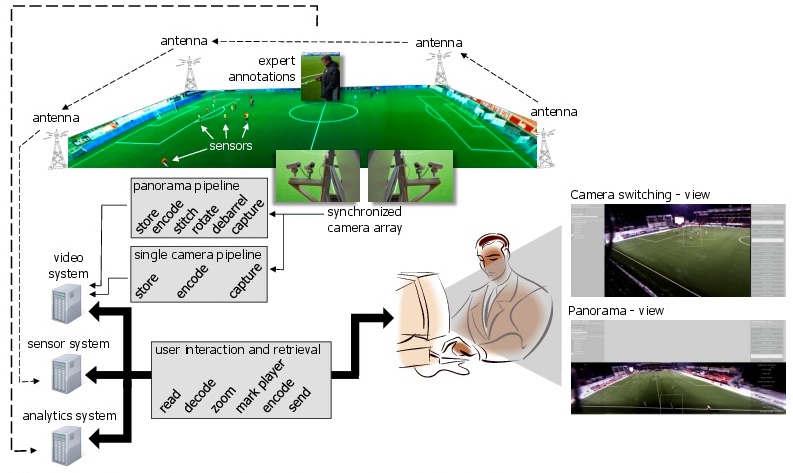}
    \caption{Overall Bagdus architecture~\cite{stensland2014bagadus}.}
    \label{fig:bagadus}
\end{figure}

Bagadus~\cite{johansen2020scalable,stensland2014bagadus,halvorsen2013bagadus,saegrov2012bagadus} is an integrated sport analysis system for real-time panorama video demonstration of athletics events. Figure~\ref{fig:bagadus} shows the overall block diagram of the Bagadus system. This system is presently used at Alfheim stadium (Troms{\o} IL) and Troms{\o}, and Ulleval stadium (Norwegian National soccer team, Oslo). A Bagadus system integrates the three main sub-systems: a sensor system (ZXY), a coach annotation system, and a video system. By utilizing these sub-systems, Bagdus installs and combines several components to stimulate the desired sports application-specific scenario. The system's modernity stems from the integration and combining of many components that allow for the autonomous demonstration of video events that are primarily dependent on data analytics and positioning sensors and are coordinated with the video system. Bagdus, for example, can extract videos of a player who is racing faster than 10 miles per hour, as well as when all of the defenders are in the 18-yard penalty box~\cite{mortensen2014automatic}. Furthermore, we may follow and pick one or more participants in the video, as well as repeat and retrieve the specific occurrences that have been marked by specialists. It used to take a long time to evaluate sports summaries and write reports. Bagdus functions as an integrated platform, handling basic processes including video synchronization automatically.  

\subsection{Player Tracking System}
There are different player tracking systems. Example includes optical-based technology, GPS and radio etc. Bagdus possibly can use any tracking. For example, at Alfheim stadium, the player tracking system imports position of the players using the ZXY Sport Tracking system~\cite{chyronhego1}. 
\subsection{Coach Annotation System}
The coach annotation system~\cite{johansen2012muithu} replaces the conventional pen-and-paper annotations, which were used previously by the coaches for soccer game annotations. Currently, with a single press of a button on Bagadus mobile app, the coaches can annotate the video very quickly. The system is based on hindsight recording, in which the end point of the event is identified and the system then takes the video prior to this mark. Furthermore, the documented occurrences are preserved in an analytics database, which can subsequently be displayed along with the related video of that specific incident. The timing of the tagged event and video must be synced here. As a result, the mobile device synchronizes its regional time with NTP machines~\cite{gaddam2018camera}. Nonetheless, it is simply a second-granularity time-sync requirement.  

\section{Individual Subjective Reports (PMSYS)}
The Player Monitoring SYStem (PMSys)~\cite{vuong2015pmsys} is a smartphone self-reporting tool that allows for the tracking of several phenotypic parameters via recurring questionnaires that players answer to via their mobile devices. PMSys supports both the IOS and Android platforms since it provides dependable and systematic reports from all team members at regular intervals. It was created as a hybrid-mobile application based on the Ionic 2+ Framework to lower the expense of multi-platform support~\cite{bendiksen2013application}. The current versions of the framework build apps that feel and look like native ones, and previous aspects and performances have been greatly reduced~\cite{pettersen2018quantified}. PMSys is now available in both Google Play and the iTunes store for Android and iOS devices. A mobile application with graphical visualization feedback provides the players with a timeline overview. The data collected via the PMSys system is also used in several follow up research that is focusing on players performance~\cite{wiik2019predicting,johansen2020scalable,riegler2016heimdallr,thambawita2020pmdata}.

\section{Distributed Fog Computing for the Big Data Generated from the Sport Analytics}

\begin{figure}
    \centering
    \includegraphics[scale=0.24]{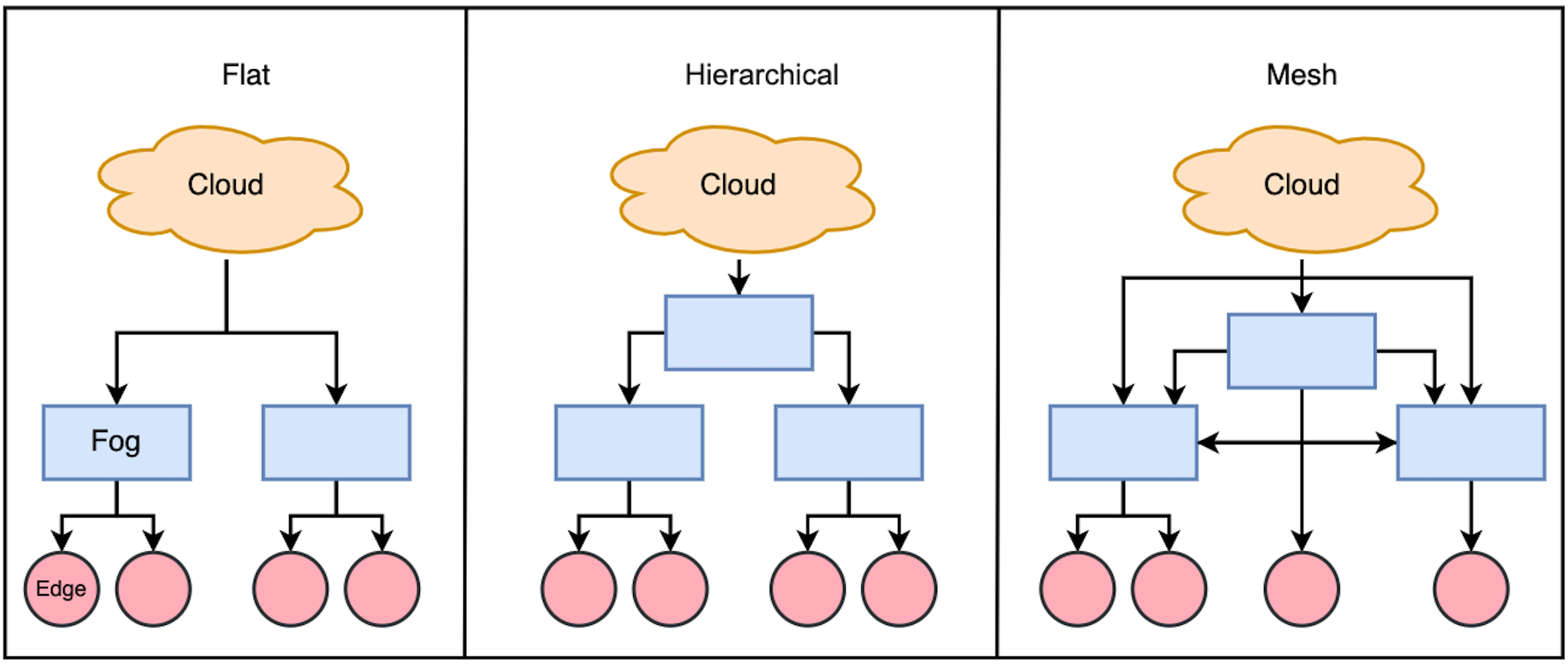}
    \caption{Interaction models present in the fog computing.}
    \label{fig:fog_computimg}
\end{figure}

Fog computing is a potential technology that tries to address current IoT paradigm challenges. Figure~\ref{fig:fog_computimg} shows the different interaction model present in the fog computing. It is appropriate in conjunction with wide-area distributed systems, with several clients at the edge of the network~\cite{donnet2007internet}. The clients may be the consumer devices (for example, smartphones, smartwatches) that are interactively used by individuals or accessories which are part of IoT (for example, cameras installed at the stadium). The client can serve both a user and an actuator that accepts control signals (for example, a fitness notification on a smartwatch), as well as a data producer (for example, heart-rate monitor from the smartwatch, video streams from cameras installed at the stadium).

Fog computing assists as a computing layer within the edge devices and the cloud in the network topology~\cite{simmhan2017big}. When opposed to fog computing, cloud computing has a higher uptime and requires a constant internet connection. Applications that use fog computing can circumvent network performance limitations in cloud computing by processing data close to edge devices. As a result, they provide a beneficial counter-balance to the prevailing paradigms. It is usually assumed that defining feature of fog computing indicates a lower network distance from the edge, however there are other network topology models~\cite{simmhan2017big}. One of the most important reasons for employing fog computing is the large amount of data created at the network's edge. IoT deployments encourage fog computing. Common web clients who solely use WWW services and content noted the expansion of Content Distribution Networks (CDN) to help those with low latency. The data from the IoT sensors is used by the player to measure and analyze several metrics in real-time~\cite{cai2017iot}. In this situation, fog computing has been operating like an inverse CDN~\cite{satyanarayanan2015edge}.

\section{Discussions}
Athlete performance analysis is a growing subject of interest in the sports industry. Wearables, electronic performance, and tracking devices are already having an impact on training and matches. The data can also be used to determine which player would be best suited to their club. As a result, analytics is referred to as the ``present and future'' of sports. Any team that does not fully utilize these will be at a competitive disadvantage. The emergence of wearables and electronic devices has expedited athlete quantification technology research and development. The German national soccer team, for example, used wearable devices to profile their players. Based on the data, coach Joachim Low made the critical substitute of Mario Götze, who scored the game-winning goal in the 2014 World Cup final in Brazil~\cite{worldcup}. It is crucial to do research on mental fatigue and post-match retrieval using mechanical workload metrics that have a logical link with neuro-muscular demands. Regardless of technical difficulties in the implemented setting, additional research on the effect of cognitive and central nervous system function, travel, sleeping behavior, feedback from the trainer, and nutritional status on Post Match Fatigue (PMF) responses would be necessary to expand the literature base. The development of mentally challenging activities with high ecological validity for soccer is critical for determining the extent to which emotional fatigue occurs in players and then tracking its time-course for recovery post-match.
 

Adapting crowdsourcing and human computation methodologies aids in the collection of real-time and interactive crowdsourced data~\cite{alonso2019practice}. During user study, spectators are required to execute monitoring tasks such as finding the players on the field, identifying ball passes etc., of a real-life soccer game. With crowdsourcing, players can acquire extensive and very complicated information in real-time. However, there are also challenges in maintaining the quality of the data source, managing the crowd and sourcing the right crowd. One of the potential solutions for addressing challenges would be transparency in data, method and research process~\cite{silberzahn2018many}.

An enormous amount of IoT data originates as observational streams in the context of distributed computing for sport analytics, or as time-series data gathered from widely distributed sensors~\cite{shukla2017riotbench,naas2017ifogstor}. These data are high-velocity data streams that are latency-sensitive that require online analytics and decision-making to give, say, health alerts of the player on/off the ground. A substantial amount of data is also generated through the video streams supplied by the stadium's cameras. As high-definition cameras become more affordable, the bandwidth necessary to drive data from the edge to the cloud can expand in volume, but network capacity remains constant~\cite{satyanarayanan2015edge}. The applications that must be approved can include real-time video analytics to appropriately record footage for future implementation.

Fog computing is prone to \textit{security attacks} such as forgery~\cite{hong2018service}, tampering~\cite{liang2020reliable}, and Jamming~\cite{tu2018security}. Additionally, there is also concern related to \textit{privacy issue} such as {`user privacy', `data privacy', `identity privacy', `usage privacy', `location privacy', and `network privacy'}~\cite{alwakeel2021overview}. Most of these attacks can be mitigated by applying countermeasures. For example, \say{Efficient encryption techniques} can be developed by designing complex encryption algorithms. Similarly, \say{decoy technique} can be used for data authentication, \say{authentication scheme} can be used for user credentials, and \say{blockchain security} can be used for network transactions~\cite{alwakeel2021overview}. In this context, we argue that fog computing can be used to move decision-making closer to the edge in order to reduce latency and analyze data in real-time. It can minimize the amount of bandwidth utilized in the core internet and limit data movement to the local network~\cite{simmhan2017big}. Furthermore, fog nodes placed near IoT devices, including end-users, can reduce both propagation delay and bandwidth utilization. 

\section{Conclusion and future perspective}
Sports analysis based on video analytics is a demanding area of research. The live data from broadcasting has been shown to be useful for match analysis and media coverage. These statistics are used by coaches, sports scientists, and the media to classify matches based on specific patterns or visual qualities that allow the categorization of match style. The motion analysis has supplied crucial information about the physical state of the players. PMSYS has been found to be advantageous to the team.

According to our study, we found that fog computing can be an excellent solution for real-time distributed processing of video streams generated by cameras and other IoT sports-related applications. As a result of the benefits of distributed fog computing and IoT, we believe it can be a possible performance evaluation solution for sports analytics. Many aspects will be improved in the future, such as the development of transparent data and advanced machine learning-based algorithms that may be used to examine performance analysis and pattern recognition in depth. Multiple sources of data can be collected in order to analyze various elements of the player and the team. More research on positioning data tracking methods, as well as fatigue and injury monitoring warrants, is necessary to offer alternative solutions. 

\bibliographystyle{IEEEtran}
\vspace{-2mm}
\bibliography{References}
\end{document}